\documentstyle[aps,floats,preprint]{revtex}
\tightenlines
\newcommand{\be}{\begin{equation}}
\newcommand{\ee}{\end{equation}}
\newcommand{\bear}{\begin{eqnarray}}
\newcommand{\eear}{\end{eqnarray}}

\begin{document}
\draft
\title{Estimating higher order perturbative coefficients using \\
Borel transform}
\author{ Kwang Sik Jeong 
and  Taekoon Lee}
\address{Department of Physics, Korea Advanced Institute of 
Science and Technology, \\
Daejon 305-701, Korea}
\maketitle
\begin{abstract}
A new method of estimating higher order perturbative coefficients
is discussed. It exploits the rapid, asymptotic growth
of perturbative coefficients and the information on the singularities
in the complex Borel plane. A comparison with other methods is 
made in several Quantum Chromodynamics (QCD) expansions.
\end{abstract}
\pacs{11.15.Bt,11.25.Db,04.25.Nx}

The ordinary perturbative expansions in weak coupling constant
in quantum field theories are 
generally asymptotic expansions, with their perturbative coefficients
growing factorially at large orders \cite{justin1}.
In practice, in many cases the 
asymptotic growth sets in quite early in perturbation, and the rapid
growth of the coefficients becomes apparent already at first few orders.

The asymptotic divergence of the perturbative coefficients implies
singularities in the Borel plane. There are three kinds of known
singularities, all on the real axis: those instanton-induced, ultraviolet 
renormalons, and infrared  renormalons.

In this paper, we show that this rapid growth of the perturbative
coefficients and the information on the singularities in the Borel
plane can be turned into a useful tool that allows us to estimate
the unknown $(N+1)$th order coefficient using the known coefficients 
up to order $N$.
The method we are going to present can be most easily understood
by working out an explicit example, 
the double-well potential in quantum mechanics.
The tunneling between the two potential wells splits the degenerate
perturbative ground states into a parity even, and an odd states,
and the average energy $E(\alpha)$ 
of the energies of these two states has the perturbative expansion of the
form
\be
E(\alpha)=-\left[\alpha+ \sum_{n=1}^{\infty} a_n \alpha^{n+1}\right]\,,
\label{energyexpansion}
\ee
where $\alpha$ denotes the canonical coupling of the model \cite{justin2}.
The Borel transform $\tilde E(b)$ of $E(\alpha)$, which has the
perturbative expansion
\be
\tilde E(b)=-\left[1+ \sum_{n=1}^{\infty} \frac{a_n}{n!} b^{n}\right]\,,
\label{borelexpansion}
\ee
is known to have multi-instanton--anti-instanton caused 
singularities at $b=2 n S_0,\, (n=1,2,3,\cdots)$, where $S_0=1/6$ is the
one-instanton action (in units of $1/\alpha$).

The nature of the singularities can in principle be determined by
doing perturbation in the background of  multi-instanton--anti-instanton 
configurations.
The closest singularity to the origin at $b=1/3$, which determines the
leading large order behavior of the expansion (\ref{energyexpansion}),
can be shown to have the following form \cite{lee1}
\be
\tilde E(b)= \frac{9}{\pi (1-3b)^2}\left[
1-\frac{53}{18}(1-3b) +O[(1-3b)^2\ln (1-3b)]\right] +
{\rm Analytic\,\, part}\,,
\ee
which comes from the instanton--anti-instanton contributions to  $E(\alpha)$.
The ``Analytic part'' denotes terms that are analytic around the
singularity.

We now consider the function $R(b) \equiv (1-3b)^2 \tilde E(b)$ as
introduced in \cite{lee3}, to control the
divergence at the singularity. $R(b)$ is bounded at $b=1/3$,
and has a very soft singularity, a suppressed logarithmic cut.
Thus the power expansion  of $R(b)$ around the origin is expected to 
better behave than that of $\tilde E(b)$. 
\begin{table}
\par
\begin{center}
{\bf Double well-potential}
\begin{tabular}{lccccc}
&$a_1$ & $a_2$ & $a_3$ & $a_4$ & $a_5$\\ \hline
Exact& 4.5&44.50&626.625&1.103 $10^4$&2.288 $10^5$ \\
Est.&6.0&33.75&600.047&1.070 $10^4$&2.266 $10^5$
\end{tabular}
\end{center}
\caption{\footnotesize The exact and the estimated coefficients
of the vacuum energy of the double-well potential.}
\label{table-instanton}
\end{table}
Ignoring the
residual logarithmic cut, we would expect the convergence radius of
the former is bounded by the second singularity at $b=2/3$,
thus effectively to become twice that of the latter.
To make the expansion of $R(b)$ better, we can take a further step
of conformally mapping the singularities except the first
one as far away as possible from the origin. This way one can hope
to have a smoother $R$ in the new plane, and a better behavior of 
the expansion.

The conformal mappings we consider in this paper are such that all the
singularities in the Borel plane except the first one are mapped
to the unit circle and the first one to a point within the  circle \cite{lee4}.
Such a mapping in this case is given by
\be
w= \frac{1-\sqrt{1-3b/2}}{1+\sqrt{1-3b/2}}
\ee
which maps the first singularity to $w_0=(\sqrt{2}-1)/(\sqrt{2}+1)=0.171$
and all others to the unit circle.
Without the residual cut-singularity, $R$ would be analytic 
on the unit disc in the $w$-plane.

Now the power expansion of  $R(b(w))$  up to $O(w^5)$ is
\bear
R&=&1 +\sum_{n=1}^{\infty} r_n w^n \nonumber \\
 &=& 1+ (-16 +2.667 a_1)w + (-120 +3.556 a_2)w^2+ (-1896.44+3.161
a_3)w^3+\nonumber \\
 && (-22544 +2.107a_4)w^4+(-254667.7+1.124a_5)w^5 +\cdots\label{inst-exp1} \\
 &=&1-4w+38.22w^2+84w^3+699.1w^4+2538.1 w^5 +\cdots\,,
\label{inst-exp2}
\eear
where the coefficient at a given order in (\ref{inst-exp1}) was calculated 
using the exact values of $a_n$, which are given in Ref. \cite{justin2}, 
up to that order less one, and
the coefficients in (\ref{inst-exp2}) were obtained with the exact values of
$a_n$.
What is interesting about this expansion is that the $a_n$-independent
constant term
in the coefficient at a given order in (\ref{inst-exp1})
is much larger than the exact value of the 
corresponding $r_n$ in (\ref{inst-exp2}).
This then implies that a good approximate estimate of $a_n$ can be
obtained by simply putting 
\be
r_n=0
\ee
in (\ref{inst-exp1}).
The estimated values for $a_n$ from this prescription
are given in Table \ref{table-instanton},
which shows improving accuracy as the order grows, and the
accuracy becomes better than 1\% at $n=5$.

Obviously, the success of this method relies on the exact value of
$r_n$ being much smaller than the constant term in the corresponding
coefficient in expansion (\ref{inst-exp1}). We may understand
this feature in the following way. We note first the coefficient $r_n$ is a 
linear combination of $a_i$'s ($i\leq n$), which are large numbers.
For the 
sake of argument, let us ignore for the moment the soft cut-singularity 
of $R$ at $w=w_0$. Then $R$ is analytic on the unit disk in $w$-plane, and so
we expect the growth of $r_n$ is fundamentally limited. To yield a small
number out of large numbers, this then suggests
a delicate cancellation occur among the large numbers in the expression for
$r_n$. However, when $r_n$'s are written as in expansion
(\ref{inst-exp1}), the cancellations are incomplete, yielding the large
constant terms. This account shows that the essential ingredients for the
success of our prescription are: (i) rapid growth of the 
perturbative coefficients, and (ii) information on the nature and 
locations of the Borel plane singularities.
One may notice the exact values of
$r_n$'s in (\ref{inst-exp2}) still grow quite rapidly. This may be
due to the residual logarithmic cut,
and the infinite number
of singularities on the unit circle.

With this example in mind, we now give the general
prescription for estimating uncalculated higher order coefficients.
Let us suppose an amplitude $A(\alpha)$ and its Borel transform
in a given theory  have the perturbative
expansion of the form
\be
A(\alpha)=\alpha +\sum_{n=1}^{\infty}a_n \alpha^{n+1}
\label{expansion3}
\ee
with $\alpha$ denoting the canonical coupling of the theory, and 
\be
\tilde A(b)=1 +\sum_{n=1}^{\infty}a_n \frac{b^n}{n!} \,.
\label{borelexpansion3}
\ee
The overall normalization of the expansion (\ref{expansion3})
was chosen such that the coefficient of the leading term is unit.
We assume  the Borel transform $\tilde A(b)$ 
has singularities on the real axis on the
$b$-plane, and the first two singularities on the positive real axis
are  at $b=b_0$, and at $b=b_1$, respectively, and the first
singularity on the negative real axis is at $b=-b_{-0}$.
We further assume the nature of the first singularity on the positive real
axis is known, and is of the form
\be
\tilde A(b) = \frac{C}{(1-b/b_0)^{1+\nu}} [1+ O(1-b/b_0)] +{\rm Analytic
\,\,part}\,,
\label{singularity2}
\ee
with $C$ a real constant and the `Analytic part' denoting terms analytic
about $b=b_0$.
We now introduce a conformal mapping
\be
w= \frac{\sqrt{1+b/b_{-0}}-\sqrt{1-b/b_{1}}}
{\sqrt{1+b/b_{-0}}+\sqrt{1-b/b_{1}}}\,,
\label{mapping2}
\ee
which maps the first singularity on the positive axis to a point within
the unit circle and all others to the circle, and the function
\be
R(b)= (1-b/b_0)^{1+\nu} \tilde A(b)\,.
\ee
We then expand $R(b(w))$ in power series in $w$-plane:
\bear
R&=&1+\sum_{n=1}^{\infty} r_n w^n \nonumber \\
&=& 1+\sum_{n=1}^{\infty} (p_n +q_n a_n) w^n \,.
\label{r-series2}
\eear
Note that $p_n, q_n$ depend linearly  on the leading coefficients 
to $a_n$ only. The estimated $n$th order coefficient is
then given by
\be
a_n =-\frac{p_n}{q_n} \,.
\label{prediction}
\ee
The fractional error of this estimate is 
\be
\delta_n = \left|\frac{r_n}{p_n}\right|\,,
\ee
and this scheme would work as long as $r_n$ is much smaller than $p_n$.
Of course, it is impossible to know the error {\it a priori}, and the
reliability of the estimated coefficients should be decided by
other circumstantial informations, for instance, as in the following examples,
by the pattern of the known terms in expansion (\ref{r-series2}), or by 
comparison with the estimated values from
using different methods. Since the expansion (\ref{r-series2}) is in general
not convergent on the whole unit disc due to the residual soft singularity,
and because of the singularities on the boundary of the disc, it is 
possible in principle for $r_n$ at a certain order to jump to a large value. 
In such a 
case the scheme could fail, but that does not seem to occur, at least,
at the orders considered in the following QCD expansions.

We now apply our scheme to the various QCD expansions, estimating
the uncalculated next higher order coefficients, and compare the 
obtained values
with the estimates from other methods. Of course, the coefficients
are renormalization scheme and scale dependent, but for simplicity,
in all of the 
following examples our consideration will be in the $\overline{\rm MS}$
scheme and at the renormalization scale  $\mu=Q$, where $Q$ 
denotes the energy scale of the problem in consideration.
The canonical coupling in 
these examples is $a(Q)\equiv \alpha_s(Q)/\pi$, where $\alpha_s(Q)$
is the strong coupling constant.

\begin{table}
\begin{center}
{\bf Bjorken polarized sum rule}
\begin{tabular}{ l| l l| l l l l| l l| l l}
$N_f$
&${d_1}^{\mathrm{ext.}}$&${d_1}^{\mathrm{est.}}$
&${d_2}^{\mathrm{ext.}}$&${d_2}^{\mathrm{est.}}$&${d_2}^{\mathrm{PMS}}$
&${d_2}^{\mathrm{ECH}}$
&${d_3}^{\mathrm{est.}}$&${d_3}^{\mathrm{PMS}}$
&${d_4}^{\mathrm{est.}}$&${d_4}^{\mathrm{PMS}}$                 \\ \hline
1    &4.25&3.86  &34.01&27.36&28.41&27.25  &302&290  &3750&2716 \\
2    &3.92&3.51  &26.94&23.17&24.09&23.11  &223&203  &2515&1696 \\
3    &3.58&3.14  &20.22&19.21&20.01&19.22  &156&130  &1565&933  \\
4    &3.25&2.73  &13.85&15.46&16.16&15.57  &101&68   &867 &396  \\
5    &2.92&2.29  &7.84 &11.90&12.59&12.19  &60 &18   &388 &56   \\
6    &2.58&1.79  &2.19 &8.48 &9.29 &9.08   &31 &-22  &95  &-115
\end{tabular}

{\bf Adler function}
\begin{tabular}{ l| l l|  l l l l| l l| l l}
$N_f$ 
&${d_1}^{\mathrm{ext.}}$&${d_1}^{\mathrm{est.}}$
&${d_2}^{\mathrm{ext.}}$&${d_2}^{\mathrm{est.}}$&${d_2}^{\mathrm{PMS}}$
&${d_2}^{\mathrm{ECH}}$
&${d_3}^{\mathrm{est.}}$&${d_3}^{\mathrm{PMS}}$
&${d_4}^{\mathrm{est.}}$&${d_4}^{\mathrm{PMS}}$                 \\ \hline
1    &1.87&3.45  &14.11&8.18&8.71&7.54  &66&75      &693&550    \\
2    &1.76&3.19  &10.16&7.19&7.55&6.57  &43&50      &391&316    \\
3    &1.64&2.90  &6.37 &6.26&6.40&5.61  &24.4&27.5  &165&151    \\
4    &1.53&2.58  &2.76 &5.37&5.27&4.68  &9.9&8.4    &8.04&49    \\
5    &1.41&2.22  &-0.69&4.50&4.16&3.77  &0.35&-7.7  &-87&2.47   \\
6    &1.30&1.80  &-3.96&3.59&3.08&2.88  &-3.48&-21  &-129&4.18
\end{tabular}
\end{center}
\caption{\footnotesize The Bjorken polarized sum rule and the Adler function:
Estimated values
along with the exact values and the PMS, ECH values when available.
}
\label{table-bjorken}
\end{table}

{\bf The Bjorken sum rule:}
Our first example is the QCD correction $\Delta$ for the Bjorken
polarized sum rule, which has the perturbative expansion
\be
\Delta(a)=a +\sum_{n=1}^{\infty} d_n a^{n+1}\,.
\ee
Incidently, this correction coincides with that of Gross-Lewellyn Smith (GLS)
sum rule at next-leading order, and differs only by the small `light-by-light'
contribution at next-next-leading order.
The first two coefficients $d_1, d_2$ are known \cite{bjpsumrule1}.
The locations of the
first singularities of the Borel transform $\tilde \Delta (b)$
and the parameter $\nu$ are given as \cite{mueller1}
\be
b_0=\frac{1}{\beta_0}, \hspace{.22in} b_1=\frac{2}{\beta_0},\hspace{.22in}
b_{-0}=\frac{1}{\beta_0},\, \hspace{.22in} \nu=
(\beta_1/\beta_0-\gamma_0)/\beta_0\,,
\ee
where $\beta_0,\beta_1$ are the first two coefficients of the
QCD $\beta$-function:
\be
\beta_0=(11-2/3 N_f)/4,\hspace{.22in}\beta_1=(102-38N_f/3)/16\,,
\ee
and $\gamma_0$ is the one-loop anomalous dimension of the twist-four
operator appearing in operator product expansion of the sum rule \cite{sv}:
\be
\gamma_0=(N_c-1/N_c)/3=8/9 \,.
\ee
Here $N_f, N_c$ denote the number of quark flavors and colors, respectively.
These are all we need to estimate the unknown next higher order coefficients.

The result is given in Table \ref{table-bjorken}, alongside of the estimates
from the  
principle of minimal sensitivity (PMS) method (and the effective charge (ECH)
method) by Kataev and Starshenko \cite{ks}.
The latter is based on the optimization of the perturbative amplitude
over the parameter space of the renormalization scheme and scale dependence.
It estimates the uncalculated higher order coefficients
by reexpanding the optimized amplitude in terms of the
coupling in the particular scheme and scale of interest.
The recent, exact, partial calculation of the next-next-next-leading order 
correction \cite{chetyrkin} indicates this method, among others
\cite{methods1}, works particularly well .

The estimates from the two methods are in good qualitative 
agreement, with
some exceptions at $d_3$.
It is remarkable that these two completely different approaches
offer such close estimates. As in PMS case, our method works best
at $N_f=3$, and the estimates become worse as $N_f$ increases.
Although the two approaches are in an overall agreement, the differences
at $d_3$ estimates are large enough to be phenomenologically significant.
For instance, the new estimate $d_3\approx 156$ at $N_f=3$, which was
used in the recent GLS sum rule analysis \cite{lee2},
results in a less 
renormalization scale dependent amplitude than with 
the PMS estimate $d_3\approx 130$.
It is instructive to see this estimate more closely. The expansion
of $R$, following (\ref{r-series2}), at $N_f=3$, reads
\bear
R&=&1+\sum_{n=1}^{\infty}r_n w^n= 1+
\sum_{n=1}^{\infty}(p_n+q_n d_n)w^n \nonumber\\
&=&1+ (-3.72+1.185 d_1) w+ (-13.488+0.702 d_2) w^2 +
(-43.257+0.277 d_3) w^3 +\cdots  \\
&=& 1+0.527 w+0.709 w^2+ \cdots
\label{bjpol-exp2}
\eear
where the last two terms in (\ref{bjpol-exp2}) are exact.
Notice that the first two $p_n$'s are large compared to the corresponding
$r_n$'s, satisfying one of the absolutely necessary conditions for 
our scheme, and the pattern of the first two known coefficients
suggests that $p_3$ is also likely to be much larger than $r_3$.
If we believe that the first two $r_n$'s have any indication on $r_3$,
it seems to be safe to assume $|r_3| \leq 2.0$. This then
leads to $ 149 \leq d_3\leq 163$.

The estimates of the next higher order coefficients $d_4$'s are
obtained using the  $d_3$ estimates.
Compared to the PMS estimates there is some significant difference at this
order. Since the obtained values depend on the estimated $d_3$,
one should note that there is another uncertainty arising 
from the error 
in $d_3$ estimate.

{\bf The  Adler function:}
The Adler function $D(a)$ for the vector 
current-current correlation function of different quark flavors
has the perturbative expansion
\be
D(a)=a +\sum_{n=1}^{\infty} d_n a^{n+1}
\ee
where the first two coefficients are known
\cite{adler1}.
Several physical observables 
can be related to this function through the dispersion relations,
and the perturbative coefficients of the former can be obtained once the 
corresponding coefficients in the Adler function are known.

The locations of the first renormalon singularities and $\nu$ in this case 
are given by \cite{mueller2}
\be
b_0=\frac{2}{\beta_0}, \hspace{.22in} b_1=\frac{3}{\beta_0},\hspace{.22in}
b_{-0}=\frac{1}{\beta_0},\, \hspace{.22in} \nu=
2\beta_1/\beta_0^2\,,
\ee
and the resulting estimates for the first coefficients
are given in Table \ref{table-bjorken}, along
with the PMS, ECH values.
Again, our method works best at $N_f=3$ for the first two coefficients,
and in this case the predicted $d_3=24.4$ is slightly smaller than
the PMS value $d_3=27.5$.

\begin{table}
\begin{center}
{\bf Quark mass}
\begin{tabular}{ l | l l l l|  l l l l| l l l| l} 
$N_f$
&${r_1}^{\mathrm{ext.}}$&${r_1}^{\mathrm{est.}}$&${r_1}^{\mathrm{res.}}$
&${r_1}^{\mathrm{lar.\;\beta_0}}$
&${r_2}^{\mathrm{ext.}}$&${r_2}^{\mathrm{est.}}$&${r_2}^{\mathrm{res.}}$
&${r_2}^{\mathrm{lar.\;\beta_0}}$
&${r_3}^{\mathrm{est.}}$&${r_3}^{\mathrm{res.}}$
&${r_3}^{\mathrm{lar.\;\beta_0}}$
&${r_4}^{\mathrm{est.}}$                                        \\ \hline
1 &9.30&7.33&6.29&12.1 &123.3&118.8&120.2&117.6 &2225&2248&1894 &51428 \\
2 &8.52&6.82&5.83&11.3 &104.8&101.3&102.3&102.9 &1762&1776&1551 &37986 \\
3 &7.74&6.28&5.34&10.5 &87.2 &84.9 &85.7 &89.2  &1357&1364&1251 &27098 \\
4 &6.96&5.71&4.83&9.8  &70.7 &69.8 &70.6 &76.5  &1008&1009&993  &18476 \\ 
5 &6.18&5.09&4.28&9.0  &55.1 &55.8 &57.2 &64.7  &712 &711 &774  &11848 \\
6 &5.40&4.43&3.76&8.2  &40.5 &43.1 &45.5 &54.0  &469 &468 &589  &6959
\end{tabular}
\end{center}
\caption{\footnotesize 
Estimated values for the quark mass expansion
along with the exact values and the  values from the residue, large $\beta_0$
methods.}
\label{table-qmass}
\end{table}

{\bf The quark mass:}
This last example is concerned with the on-shell quark mass.
The on-shell quark mass and the $\overline {\rm MS}$ mass are related by
\be
\frac{m_{\rm OS}}{m_{\overline{\rm MS}}}=1+ \frac{4}{3}\left[
a + \sum_{n=1}^{\infty} r_n a^{n+1}\right]\,,
\ee
and the first two coefficients are known \cite{qmass1}.
The relevant parameters in this case are \cite{qmass3}
\be
b_0=\frac{1}{2\beta_0}, \hspace{.22in} b_1=\frac{3}{2\beta_0},\hspace{.22in}
b_{-0}=\frac{1}{\beta_0},\, \hspace{.22in} \nu=
\beta_1/2\beta_0^2\,,
\ee
and the estimated first coefficients are in Table \ref{table-qmass},
along with the values from the  residue based
method of Pineda \cite{pineda} (also, for comparison, those of
large $\beta_0$ approximation \cite{largebeta}),
which is known to work well in this case.
The latter method relies on the
rapid convergence of the perturbative calculation of the renormalon 
residue, and the expansion of the Borel transform about the first renormalon
\cite{lee3,lee4}. 
The agreement between  these two approaches at $r_3$ estimates
 is remarkable.

To conclude the paper,
we presented a new method of estimating higher order unknown coefficients.
It is based on the two generic features of
the asymptotic expansions,
the rapid growth of the coefficients and
the Borel plane singularities.
It can provide independent estimates for the
yet unknown coefficients, which may prove useful in physical analysis.

\acknowledgments
T.L. is thankful to Gorazd Cvetic for useful discussions and comments.
This work was supported by the BK21 Core Program.

\vspace{1in}
{\bf Postscript}: After submission of our paper we were informed
that in the published version \cite{cap-fischer} of hep-ph/9811367
Borel transform with a conformal mapping was tried on a model function
to estimate higher order coefficients. Unlike our method, however, it
did not remove the first Borel singularity, and because of that
it reached to a conclusion substantially different from ours,
that the method can work only for
large orders (say, $N\sim 7$)
and is not applicable for low orders (N$\sim$ 3) relevant
for present QCD perturbations.



\end{document}